\documentclass{article}

\usepackage[utf8]{inputenc} 
\usepackage{amsmath,amssymb}
\usepackage{graphicx} 
\usepackage{hyperref} 
\usepackage{bm}

\title{On the Lagrangian and kinetics of a massive and charged spin-one vector boson}

\author{
  Eckart Marsch\\
  Retired Research Scientist, 
   Johann-Fleck-Stra{\ss}e, 24106 Kiel, Germany\\
  \and
  Yasuhito Narita \\
   Institut f\"{u}r Theoretische Physik, 
   Technische Universit\"{a}t Braunschweig,\\ 
   Menselssohnstr. 3, 38106 Braunschweig, Germany \\
   also at Max Planck Institute for Solar System Research, \\
   Justus-von-Liebig-Weg 3, 37077 G\"ottingen, Germany \\
  \texttt{y.narita@tu-braunschweig.de} \\
}

\date{} %

\begin{document}
\maketitle

\begin{abstract}
A new kinetic equation and Lagrangian of a massive and charged spin-one vector boson is outlined in this paper. Those quantities are derived by exploiting directly the Lorentz group generators including the boson spin. Thereby, the chiral symmetry of the Lorentz transformation is guiding the algebraic calculations. Two associated 4\,x\,4 spin and helicity matrices can be defined, the sum of which give the rotation operator that yields in the presence of an electromagnetic field the boson-spin coupling to the magnetic field. The four polarization vectors of the boson field are calculated and discussed. Moreover, the current density is derived and used to reveal the bosonic nature of the kinetic equations. The energetic stability of the vector boson quantum field is shown. A Pauli equation for the vector boson is derived.
\end{abstract}

\section{Introduction}

Massless vector bosons without charge, such as for example the prominent photons in quantum electrodynamics, have traditionally been described by Maxwell's equations in either classical \cite{jackson} or quantized \cite{schwartz} form. In this paper we address the question whether it is possible to make the spin-one property of a charged and massive vector boson to appear explicitly in its kinetic equation, like it is the case for the spin one-half of a fermion such as the electron described by the Dirac equation \cite{dirac1928}. 

The search for vector bosons beyond the Standard Model (SM) has gained significant attention due to their potential role to mediate parity violation and act as dark matter candidates \cite{bertone2005,bertone2018,gaul2026}. The description of the SM massive vector bosons rests basically on the Proca equation  \cite{proca1936,proca1938,proca1939,proca2006} (Maxwell equation generalized  by a mass term). The present approach differs essentially from the Proca equation as there is no constraint such as the Lorentz condition, $\partial_\mu A^\mu=0$, which is placed on the four-vector gauge field $A^\mu$ and stems from the constructed Lagrangian \cite{schwartz}. We shall assume here that the mass of the four-vector field is given, which we name $\phi^\mu=(\phi_0, \bm{\phi})$ and abbreviate concisely as $\phi$. Concerning relations to massless gauge fields, we only consider coupling to the electromagnetic field. Our Lagrangian is physically motivated and derived by means of the generators of the Lorentz group.

Historically, Dirac achieved his goal with the help of matrix algebra, by linearization of the Lorentz-invariant d'Alembert differential operator that is of second order in space-time. This procedure resulted in the famous $4 \times 4$ matrices named after him, which include the four fermion degrees of a freedom for spin up and down as well as particle and antiparticle. His approach was, according to modern understanding \cite{schwartz}, founded on the spinor representation $su(2) \oplus su(2)$ of the Lorentz algebra. Yet, a similar equation was not invented for a charged and massive vector boson of spin one, which is unlike the photon not its own antiparticle. Here we make a first attempt to do this by exploiting the $SO(3,1)$ representation symmetry of the Lorentz group for vectors.

Therefore, we exploit the two $4\times4$-matrix three-vectors of rotation and boost, as they occur in the Lorentz transformation of four-vectors in Minkowski space-time, in order to develop a second-order differential equation for vector fields which can be massive and charged. Two related $4\times4$ spin matrices can be defined and derived directly from the Lorentz-group generators. They reflect the chiral symmetry and correspond to spin one. Lorentz invariance of the related eigen or polarization vectors requires two of those to be transverse, but includes as well the single longitudinal polarization. These three degrees of freedom are a direct consequence of the spin-one property of the vector-field $\phi$. The introduced spin or related helicity matrices are found to shape explicitly the kinetic equation of the vector boson, and they yield in the presence of an electromagnetic field also the spin-coupling to a magnetic field. 

Furthermore, they appear in the Lagrangian of the vector field and its Noether current, which both are derived in this paper. The bosonic nature of the quantum vector field is established via the current density by the requirement that the involved creation and annihilation operators must be connected by commutators. The Hamiltonian or energy density as being derived from the energy-momentum tensor of the vector field are found to be bounded from below after quantization, and thus the physical meaning of the Lagrangian presented is clear and credible.  

To avoid misunderstandings at the outset, here we do certainly not want to rival the established modern quantum field theory \cite{schwartz,kaku}, in which for instant the dynamics of the two $W^{\pm}$ bosons are described by massive, electrically charged spin-one vector fields. They are governed by a non-Abelian Yang-Mills \cite{yangmills} Lagrangian density coupled to the Higgs sector \cite{higgs}, whereby their mass term arises from the spontaneous symmetry breaking of the Higgs vacuum. Furthermore, the non-Abelian nature of the involved $SU(2)$ gauge group necessitates nonlinear self-interactions (cubic and quartic gauge coupling terms) in that Lagrangian, a feature that is for example absent in the linear Dirac equation for leptons \cite{weinberg}, as well as in the basic model of a single spin-one boson presented here.

%%
%% sec. 2, Lorentz group
%%
\section{The generators of the Lorentz group}\label{sec2}

It is well known since the pioneering work of various authors \cite{wigner,bargman,joos} that the Lie algebra for the Lorentz group (LG) can be decomposed into two commuting and thus independent sub-algebras, which define the generators of the irreducible $SU(2) \otimes SU(2)$ representation of the LG. These three-vector generators can be written as tensors or $4\times4$ matrices in Minkowski space-time \cite{jackson}. 
Here we make use of the hermitian rotation operator $\mathbf{J}=(J_\mathrm{x}, J_\mathrm{y}, J_\mathrm{z})$ and the anti-hermitian boost operator $\mathbf{K}= (K_\mathrm{x}, K_\mathrm{y}, K_\mathrm{z})$. The components of these three-vector generators of the vectorial LG are $4\times4$-matrices, which are quoted explicitly for completeness and reference in the Appendix. 

According to their definitions, the rotation and boost operators obey the linked three-vector equations of the Lorentz algebra, which can be written concisely as
\begin{equation}
	\label{eq:1}
	\mathbf{J}\times\mathbf{J}=\mathrm{i}\mathbf{J}, \;\; \mathbf{K}\times\mathbf{K}=
	-\mathrm{i}\mathbf{J},  \;\; \mathbf{J}\times\mathbf{K}=\mathbf{K}\times\mathbf{J}=
	\mathrm{i}\mathbf{K}.
\end{equation}
The symbol $\times$ denotes as usually the three-vector cross product. We define in addition to the above generator the diagonal metric matrix $\Delta=\mathrm{diag}[1,-1,-1,-1]$, which corresponds to the Euclidian metric of Minkowski space. Then one finds that $[\mathbf{J}, \Delta]=0$, and that $\{\mathbf{K}, \Delta\}=0$, i.e, the rotation commutes and the boost anticommutes with Delta. Of course, $\Delta^2={\sf{1}_4}$.

%%
%% sec. 3, vector boson
%%
\section{Kinetic equation of the massive vector boson}\label{sec3}

Recently, Marsch and Narita \cite{mana2023} derived an extended Dirac equation for a fermion of spin one-half on the basis of the above vector representation of the Lorentz group. In this section we continue their approach to obtain a kinetic equation for the massive spin-one vector boson. Starting point is the relativistic energy for any massive particle, which goes with the momentum squared in the dispersion relation and is given by
\begin{equation}
	\label{eq:2}
	E^2 - \mathbf{p}^2 = M^2 = P^\mu P_\mu.
\end{equation}
This so-called mass-shell condition for any free particle of mass $M$, energy $E$ and momentum $\mathbf{p}$, and is just the first Casimir operator of the Lorentz group. According to relativistic quantum mechanics and quantum field theory (QFT) \cite{kaku,schwartz}, the covariant four-momentum quantum operator is given by  the temporal and spatial derivatives, 
%%
%% Eq. 3
%%
\begin{equation}
	\label{eq:3}
	P_\mu = (E, -\mathbf{p}) 
            = \mathrm{i}\partial_\mu 
            = \mathrm{i} ({\partial_t}, \partial_\mathbf{x} )
            = \mathrm{i} (\frac{\partial}{\partial t} ,
                \frac{\partial}{\partial \mathbf{x}}).
\end{equation}
It operates on the vector quantum field $\phi(t,\mathbf{x})$ of interest here or on its adjoint field $\bar{\phi}(t,\mathbf{x})$, which is defined as $\bar{\phi}=(\Delta \phi)^\dagger$, such that the scalar product $\bar{\phi}\phi$ is ensured to be Lorentz invariant. We abbreviate the contravariant space time coordinates $x^\mu=(t,\mathbf{x})$ by $x$ and use conveniently units of $c=1$ and $\hbar=1$. The above differential operator $P_\mu$ is going to be used later, when we discuss the desired kinetic equation. If we multiply the relativistic dispersion relation (\ref{eq:2}) with the 4-dimensional unit matrix, we obtain \cite{mana2023} by means of (\ref{eq:1}) and the related results in the Appendix the novel algebraic relation
\begin{equation}
\label{eq:4}
E^2{\sf{1}_4} + (\mathbf{K}\cdot\mathbf{p})^2 - (\mathbf{J}\cdot\mathbf{p})^2 = E^2{\sf{1}_4} - (\mathbf{J}\cdot\mathbf{p} + \mathbf{K}\cdot\mathbf{p})(\mathbf{J}\cdot\mathbf{p} - \mathbf{K}\cdot\mathbf{p}) = M^2{\sf{1}_4}.
\end{equation}
The symbol $\cdot$ denotes as usually the three-vector scalar product. When using these matrix three-vectors, which are also quoted in the Appendix, we obtain with the definition $\mathbf{L}^\pm = \mathbf{J} \pm \mathbf{K}$ three new purely imaginary $4\times4$ matrices, which we can write as the $\mathbf{L}^\pm$ matrix three-vector in component form
\begin{equation}
\label{eq:5}
L^\pm_\mathrm{x}= \mathrm{i} \left(
\begin{array}{cccc}
0 & \mp 1 & 0 & 0 \\
\mp 1 & 0 & 0 & 0 \\
0 & 0 & 0 & -1 \\
0 & 0 & 1 & 0 \\
\end{array}
\right),
L^\pm_\mathrm{y}= \mathrm{i} \left(
\begin{array}{cccc}
0 & 0 & \mp 1 & 0 \\
0 & 0 & 0 &  1 \\
\mp 1 & 0 & 0 & 0 \\
0 & - 1 & 0 & 0 \\
\end{array}
\right),
L^\pm_\mathrm{z}= \mathrm{i} \left(
\begin{array}{cccc}
0  & 0 & 0 & \mp 1 \\
0 & 0 & -1 & 0 \\
0 & 1 & 0 & 0 \\
\mp 1 & 0 & 0 & 0 \\
\end{array}
\right), 
\end{equation}	
which play a key role in what follows. Then we can rewrite (\ref{eq:4}) in the two equivalent forms
\begin{equation}
\label{eq:6}
E^2{\sf{1}_4} - (\mathbf{L^+}\cdot\mathbf{p})(\mathbf{L^-}\cdot\mathbf{p}) = E^2{\sf{1}_4} - (\mathbf{L^-}\cdot\mathbf{p})(\mathbf{L^+}\cdot\mathbf{p}) = M^2{\sf{1}_4}.
\end{equation}
By insertion of the quantum four-momentum operator (\ref{eq:3}) we obtain a second-order wave equation for the vector quantum field $\phi(t,\mathbf{x})$ and its adjoint field $\bar{\phi}(t,\mathbf{x})$. The two similar field equations read
\begin{equation}
\label{eq:7}
\begin{array}{c}
\left(\frac{\partial^2}{\partial t^2} + M^2 \right)\phi(t,\mathbf{x})  = (\mathbf{L}^+ \cdot \frac{\partial}{\partial \mathbf{x}} )
(\mathbf{L}^- \cdot \frac{\partial}{\partial \mathbf{x}} )
\phi(t,\mathbf{x}). \\
\\
\left(\frac{\partial^2}{\partial t^2} + M^2 \right)\bar{\phi}(t,\mathbf{x})  = \frac{\partial}{\partial \mathbf{x}} \frac{\partial}{\partial \mathbf{x}} \bar{\phi}(t,\mathbf{x}) : \mathbf{L}^- \mathbf{L}^+. 
\end{array}
\end{equation}
Here $\mathbf{L}^\pm$ appear symmetrically in the sign superscript. We stress that the scalar products on the right-hand sides commute with each other. The three-vector $\mathbf{L}^\pm$ has the important properties, $\Delta \mathbf{L}^\pm \Delta = \mathbf{L}^\mp$, and $(\mathbf{L}^\pm)^\dagger = \mathbf{L}^\mp$. Let us consider the differential operators on the right-hand side of (\ref{eq:7}) more explicitly. We obtain
\begin{equation}
\label{eq:71}
\mathbf{L}^\pm \cdot \frac{\partial \,\phi }{\partial \mathbf{x}} = \mathrm{i} \left(
\begin{array}{cccc}
0 & \mp \frac{\partial}{\partial x} & \mp \frac{\partial}{\partial y} & \mp \frac{\partial}{\partial z}  \\
\mp \frac{\partial}{\partial x} & 0 & -\frac{\partial}{\partial z}  &  \frac{\partial}{\partial y} \\
\mp \frac{\partial}{\partial y} &  \frac{\partial}{\partial z}  & 0 & -\frac{\partial}{\partial x} \\
\mp \frac{\partial}{\partial z}  & - \frac{\partial}{\partial y} & \frac{\partial}{\partial x} & 0 \\
\end{array}
\right) 
\left(
\begin{array}{c}
\phi_0(t,\mathbf{x}) \\
\phi_1(t,\mathbf{x}) \\
\phi_2(t,\mathbf{x}) \\
\phi_3(t,\mathbf{x}) \\
\end{array}
\right).
\end{equation}	
\newline
When multiplying the right-hand term of (\ref{eq:7}) out, we just retain the Laplacian differential operator. To show this let us, with the definition 
$\phi = (\phi_0, \bm{\phi})^T$, abbreviate first (\ref{eq:71}) as follows
\begin{equation}
\label{eq:72}
\mathbf{L}^\pm \cdot \frac{\partial }{\partial \mathbf{x}}
\left( 
\begin{array}{c}
\phi_0 \\
\bm{\phi} \\
\end{array}	
\right) = \mathrm{i} \left(
\begin{array}{c}
\mp \nabla \cdot \bm{\phi} \\
\mp \nabla  \phi_0 + \nabla \times \bm{\phi}  \\
\end{array}
\right).
\end{equation}	 
When operating with (\ref{eq:71}) on this last expression a second time, we get
\begin{equation}
\label{eq:73}
(\mathbf{L}^\pm \cdot \frac{\partial }{\partial \mathbf{x}})
(\mathbf{L}^\mp \cdot \frac{\partial }{\partial \mathbf{x}})
\left( 
\begin{array}{c}
\phi_0 \\
\bm{\phi} \\
\end{array}	
\right) =  \left(
\begin{array}{c}
\nabla^2 \phi_0 \mp \nabla \cdot (\nabla \times \bm{\phi}) \\
\nabla (\nabla \cdot \bm{\phi} ) - \nabla \times \nabla \times \bm{\phi}  \\
\end{array}
\right) = \nabla^2 \left(
\begin{array}{c}
\phi_0 \\
\bm{\phi} \\
\end{array}
\right).
\end{equation}	 

Thus equation (\ref{eq:7}) gives nothing but the Klein-Gordon equation for each of the four independent degrees of freedom of the four-vector quantum field. We may assume that $\phi$ is real and an element of the Minkowski space, or if charged it may be complex and an element of $\mathbb{C}^4$. As $\mathbf{L}^\pm$ is purely imaginary, the right-hand side of (\ref{eq:7}) is real, and thus the whole equation is real as well. However, to make the wave equations (\ref{eq:7}) hermitian, we make use of the $3 \times 3$ tensor $\underline{\underline{\mathrm{L}}}$ that is defined as 
\begin{equation}
\label{eq:74}
\underline{\underline{\mathrm{L}}} = \frac{1}{2}(
\mathbf{L^+}\mathbf{L^-} + \mathbf{L^-}\mathbf{L^+} ), 
\end{equation}
and obeys $\underline{\underline{\mathrm{L}}}^\dagger = \underline{\underline{\mathrm{L}}}$ and $\Delta \underline{\underline{\mathrm{L}}} \Delta = \underline{\underline{\mathrm{L}}}$. Thus we obtain the more symmetric expressions
\begin{equation}
\label{eq:75}
\begin{array}{c}
\left(\frac{\partial^2}{\partial t^2} + M^2 \right)\phi(t,\mathbf{x})  = \underline{\underline{\mathrm{L}}} : \frac{\partial}{\partial \mathbf{x}}  \frac{\partial}{\partial \mathbf{x}} \phi(t,\mathbf{x}). \\
\\
\left(\frac{\partial^2}{\partial t^2} + M^2 \right)\bar{\phi}(t,\mathbf{x})  = \frac{\partial}{\partial \mathbf{x}} \frac{\partial}{\partial \mathbf{x}} \bar{\phi}(t,\mathbf{x}) : \underline{\underline{\mathrm{L}}}. 
\end{array}
\end{equation}

Here $\mathbf{L}^\pm$ relates to the chiral kinetic helicity of the vector boson. This becomes more obvious when we introduce the related hermitian and purely imaginary Sigma matrices (akin to the Pauli matrices), defined as $\bm{\Sigma}^\pm = -\mathbf{L}^\pm \Delta $. They are given explicitly in the Appendix. Replacing  $\mathbf{L}^\pm$ in (\ref{eq:7}) by the Sigma matrix yields the two equivalent real chiral equations 
\begin{equation}
\label{eq:8}
\left(\frac{\partial^2}{\partial t^2} + M^2 \right)\phi(t,\mathbf{x})  = (\bm{\Sigma}^\pm \cdot \frac{\partial}{\partial \mathbf{x}} )^2
\phi(t,\mathbf{x}). 
\end{equation}
In Fourier space the right-hand side is the square of the chiral helicity operator that is defined as $\bm{\Sigma}^\pm \cdot \mathbf{p}$, whose square is $p^2{\sf{1}_4}$, and whose eigenvalues are $\pm 1$. These equations have recently  been analysed in detail by Marsch and Narita \cite{mana2026}. In order to make the wave equations (\ref{eq:8}) chirally symmetric, we define another $3 \times 3$ tensor $\underline{\underline{\Sigma}}$ that reads 
\begin{equation}
\label{eq:76}
\underline{\underline{\Sigma}} = \frac{1}{2}(
\bm{\Sigma}^+ \bm{\Sigma}^+ +  \bm{\Sigma}^- \bm{\Sigma}^-)
= \underline{\underline{\mathrm{L}}} =
\left( 
\begin{array}{ccc}
{\sf{1}}_4  &   \mathrm{i}J_z & -\mathrm{i}J_y \\
-\mathrm{i} J_z &      {\sf{1}}_4 & \mathrm{i} J_x \\
\mathrm{i} J_y & -\mathrm{i} J_x & {\sf{1}}_4     \\
\end{array}
\right). 
\end{equation}
Therewith we obtain again, by adding up the two chiral equations (\ref{eq:8}), the same wave equations as in (\ref{eq:75}). Note that the tensor elements here are themselves $4 \times 4$ matrices given by the matrices of the components of the LG rotation operator presented in the Appendix in equation  
(\ref{eq:A1}).

It should be mentioned in this context that making use of the Sigma matrices offers the possibility to derive also \cite{mana2020} the linear Dirac equation on the basis of the vector representation of the Lorentz group, and thus to provide an interesting view on the chiral symmetry of fermions coming in $SU(2)$-doublets. Yet, we shall continue here with the new quadratic version (\ref{eq:7}) and (\ref{eq:75}) of the kinetic equation for bosons following directly from (\ref{eq:4}).

%%
%% sec. 4, Lagrangian
%%
\section{The Lagrangian of the vector field}\label{sec4}

As compared with the scalar spin-zero quantum field the present spin-one vector field is more complex, as it contains explicitly the $\mathbf{L}^\pm$ operator that indicates the spin-one property of that field. The associated Lagrangian density can be constructed such that it delivers the equations of motion and is hermitian, i.e. $\mathcal{L}^\dagger=\mathcal{L}$. Thus it is given by the expression
\begin{equation}
	\label{eq:23}
	\mathcal{L}   
	=  \frac{\partial\bar{\phi}}{\partial t} \frac{\partial\phi}{\partial t} - \frac{1}{2} \left( (\frac{\partial}{\partial \mathbf{x}} \bar{\phi} \cdot \mathbf{L}^+) (\mathbf{L}^- \cdot \frac{\partial}{\partial \mathbf{x}}\phi) + (\frac{\partial}{\partial \mathbf{x}} \bar{\phi} \cdot \mathbf{L}^-) (\mathbf{L}^+ \cdot \frac{\partial}{\partial \mathbf{x}}\phi) \right) - M^2 \bar{\phi}\phi
\end{equation}
Obviously, the mass term is Lorentz invariant. The time-derivate and spatial-derivative terms are hermitian conjugated. They will be shown below to be combined into a Lorentz-invariant form as well. In this Lagrangian $\mathcal{L}(\phi,\bar{\phi},\partial_t \phi,\partial_t \bar{\phi}, \partial_\mathbf{x} \phi,\partial_\mathbf{x} \bar{\phi})$ the field $\phi$ and its adjoint $\bar{\phi}$ and their derivatives are treated equally. The equation of motion for $\phi$ results from the Euler-Lagrange equation,
\begin{equation}
	\label{eq:24}
	\frac{\partial \mathcal{L}}{\partial \phi} - \partial_t \frac{\partial \mathcal{L}}{\partial (\partial_t \phi)} - \partial_\mathbf{x} \cdot \frac{\partial \mathcal{L}}{\partial (\partial_\mathbf{x} \phi)} = 0.
\end{equation}
Similarly, the one for the adjoint field is obtained:
\begin{equation}
	\label{eq:25}
	\frac{\partial \mathcal{L}}{\partial \bar{\phi}} - \partial_t \frac{\partial \mathcal{L}}{\partial (\partial_t \bar{\phi})}  - \partial_\mathbf{x} \cdot \frac{\partial \mathcal{L}}{\partial (\partial_\mathbf{x} \bar{\phi})} = 0.
\end{equation}
The application of these equations on $\mathcal{L}$ reproduces the original equations of motion as presented already in (\ref{eq:75}). We note that the Lagrangian (\ref{eq:23}) can also be written in another form. For that purpose we use the previous relation (\ref{eq:76}) and insert it into (\ref{eq:23}). This yields
%%
%% Eq. 18
%%
\begin{equation}
	\label{eq:251}
	\mathcal{L}  
	=  \frac{\partial\bar{\phi}}{\partial t} 
           \frac{\partial\phi}{\partial t} 
            - \left( \frac{\partial}{\partial \mathbf{x}} \bar{\phi} 
                 \cdot 
                \underline{\underline{\mathrm{L}}} 
                 \cdot 
                \frac{\partial}{\partial \mathbf{x}} \phi) \right)  
            - M^2 \bar{\phi}\phi.
\end{equation}
We can further rewrite the kinetic part of the Lagrangian in another way and express it in terms of the tensor $L^{\mu\nu}$ with its four-matrix elements given in terms of the rotation operator as quoted in the Appendix in (\ref{eq:A1}). After some algebra, an enlightening result is obtained,
\begin{equation}
	\label{eq:257}
	\mathcal{L}_k =  \partial_\mu \bar{\phi} \,L^{\mu\nu} \,  \partial_\nu \phi, \;\; 
	L^{\mu\nu} = g^{\mu\nu} + J^{\mu\nu}, 
\end{equation}
which can be written as the sum of the metric tensor in  Minkowski space and the rotation tensor that is given by the hermitian tensor operator
\begin{equation}
	\label{eq:259}
	J^{\mu\nu}= 
	\left(
	\begin{array}{cccc}
		0 & 0 & 0 & 0 \\
		0 & 0 & -\mathrm{i}J_z & \mathrm{i}J_y \\
		0 & \mathrm{i} J_z & 0 & -\mathrm{i} J_x \\
		0 & -\mathrm{i} J_y & \mathrm{i} J_x & 0 \\
	\end{array} 
	\right). 
\end{equation}	
Note that the elements $J^{\mu\,0}$ and $J^{0\,\nu}$ are zero, and thus there are no connections to the time derivatives in (\ref{eq:259}), as is obvious from (\ref{eq:23}). We recall that the rotation matrix three-vector $\mathbf{J}$ is a generator of the Lorentz group as discussed in Section \ref{sec2}. It acts on $\phi$, and its three vector components are the $4\times4$ matrices of (\ref{eq:A1}), which in their spatial elements contain the $3\times3$ matrices of spin one $\mathbf{S}$ of the vector boson. Finally, we can write the Lagrangian in the compact Lorentz-invariant form 
\begin{equation}
	\label{eq:260}
	\mathcal{L} =	\partial_\nu \bar{\phi} \partial^\nu \phi -
	\partial_\mu \bar{\phi} \,J^{\mu\nu} \, \partial_\nu \phi - 
	M^2 \bar{\phi}\phi. 
\end{equation}
For vanishing spin term this Lagrangian is just the one describing four scalar spin-zero bosons assembled artificially into a four-vector with entirely independent components. The application of the Euler-Lagrange equations then reproduces the Klein-Gordon equation for each of them. However, according to the nature of the $SO(3,1)$ group symmetry, the present Lagrangian splits naturally into scalar and three-vector sections of $\phi$. This is a major difference to the Proca Lagrangian, which does nowhere contain an explicit spin term. The resulting scalar section has the simplest Lagrangian reading
\begin{equation}
	\label{eq:261}
	\mathcal{L}_0 = \partial_\nu \phi^*_0 \partial^\nu \phi_0 - M^2 \phi^*_0 \phi_0. 
\end{equation}
By disregarding it, eliminates one component of the four-vector field, corresponding to the spin-zero scalar field of the representation $SO(3,1)$ of the proper Lorentz group \cite{schwartz}. In the Proca theory this elimination is enforced by the Lorentz condition, which would imply in our case that $\partial_\mu \phi^\mu=0$. The present approach does not delivers such constraint, but the sector separation is a natural consequence of the symmetry. Namely, for the three-vector section we obtain formally an additional spin term and a Lagrangian reading
\begin{equation}
	\label{eq:262}
	\mathcal{L}_3 =  \partial_\nu \bm{\phi}^\dagger \partial^\nu \bm{\phi} - (\frac{\partial}{\partial \mathbf{x}} \bm{\phi}^\dagger \cdot \underline{\underline{S}}  \cdot \frac{\partial}{\partial \mathbf{x}}\bm{\phi}) - M^2 \bm{\phi}^\dagger \bm{\phi}. 
\end{equation}
In analogy to equations (\ref{eq:75}) and (\ref{eq:76})  we define the spin-one $3 \times 3 $ tensor
\begin{equation}
	\label{eq:777}
	\underline{\underline{S}} =
	\left( 
	\begin{array}{ccc}
		0 &   -\mathrm{i}S_z & \mathrm{i}S_y \\
		\mathrm{i} S_z &  0  & -\mathrm{i} S_x \\
		-\mathrm{i} S_y & \mathrm{i} S_x &  0 \\
	\end{array}
	\right). 
\end{equation}
The associated spin-one matrix vector reads 
\begin{equation}
	\label{eq:263}
	S_\mathrm{x}= \mathrm{i} \left(
	\begin{array}{ccc}
		0 & 0 & 0 \\
		0 & 0 &-1 \\
		0 & 1 & 0 \\
	\end{array}
	\right), \;
	S_\mathrm{y}= \mathrm{i} \left(
	\begin{array}{ccc}
		0 & 0 & 1 \\
		0 & 0 & 0 \\
		-1 & 0 & 0 \\
	\end{array}
	\right), \;
	S_\mathrm{z}= \mathrm{i} \left(
	\begin{array}{ccc}
		0 &-1 & 0 \\
		1 & 0 & 0 \\
		0 & 0 & 0 \\
	\end{array}
	\right). 
\end{equation}	
However, when multiplying the spin term in (\ref{eq:262}) out, it is found to vanish identically for a vector field $\bm{\phi}$ that is real, namely one obtains in the equation of motion that
\begin{equation}
	\label{eq:263}
	\underline{\underline{S}} : \frac{\partial}{\partial \mathbf{x}} \frac{\partial}{\partial \mathbf{x}}\bm{\phi} = \mathbf{S} \cdot (\nabla \times \nabla) \bm{\phi} = 0.
\end{equation}
So, the spin has no direct impact on the polarization of the free real vector-boson field, but it reemerges vitally and becomes important if one replaces the spatial derivatives by their covariant derivatives. This is shown subsequently and gives the coupling to the magnetic field in the presence of an electromagnetic four-vector potential.

%%
%% sec. 5, helicity
%%
\section{Eigenvectors of the helicity operator}\label{sec5}

Moreover, when it comes to determining the possible polarization vectors of the field, the kinetic equations (\ref{eq:7}) and (\ref{eq:75}) matter. For the free field we can then chose the three eigenvectors of the spin, or more adequately of the spin helicity matrix, which is defined as $H(\hat{\mathbf{p}}) = \mathbf{S} \cdot \hat{\mathbf{p}}$. Here we used the particle momentum vector normalized to unity, i.e., $\hat{\mathbf{p}}=(x,y,z)^T$, with $x^2+y^2+z^2=1$. The associated helicity matrix reads explicitly 
\begin{equation}
	\label{eq:9}
	H(\hat{\mathbf{p}}) = \mathrm{i} \left(
	\begin{array}{ccc}
		0 & -z & y \\
		z & 0 & -x \\
		-y & x & 0 \\
	\end{array}
	\right). 
\end{equation}	
and relates to the rotation of three-vectors in $\mathbb{R}^3$. It has in real space three orthogonal eigenvectors. Thus for the spin-one case, one also must consider the longitudinal polarization. Obviously, the eigenvector of $H(\hat{\mathbf{p}})$ with eigenvalue zero reads $\bm{\chi}_0 = \left(x,y,z\right)^T$. The orthogonal eigenvectors of $H(\hat{\mathbf{p}})$ with eigenvalues $\pm1$ read
\begin{equation}
	\label{eq:12}
	\bm{\chi}_\pm = \frac{1}{\sqrt{2(1-y^2)}} \left(
	\mp \mathrm{i} z - xy, \, 1-y^2, \, \pm \mathrm{i} x - yz \right)^T.
\end{equation} 
They are perpendicular to the eigenvector $\bm{\chi}_0$, which is equal to the momentum three-vector $\hat{\mathbf{p}}$. Valid for any momentum orientation, these three eigen-vectors of the spin helicity-polarization, named $\bm{\chi}_0$, and $\bm{\chi}_\pm$, form an orthonormal basis in the real space $\texttt{R}^3$ for the 3-vector component $\bm{\phi}$.
It turns out that these solutions may be constrained without loss of generality to be two-dimensional, i.e., by putting $y=0$, which implies that $x^2+z^2=1$. Then the related four-momentum reads, $p^\mu=(E(p), px, 0, pz)^T$. 

The above two eigenvectors of $H(\hat{\mathbf{p}})$ with eigenvalues $\pm1$ permit us readily to construct two physically admissible four-vectors in Minkowski spacetime, which can be written as $\epsilon^\mu_\pm=(0,\bm{\chi}_\pm)$. These solutions are
\begin{equation}
	\label{eq:13}
	\epsilon_\pm = \frac{1}{\sqrt{2}} 
	\left(0, \mp \mathrm{i} z, 1, \pm \mathrm{i} x \right)^T, \;\;
	\bar{\epsilon}_\pm = \frac{1}{\sqrt{2}} 
	\left(0, \mp \mathrm{i} z, -1, \pm \mathrm{i} x \right).
\end{equation}
But let us return to the more general solutions based on (\ref{eq:12}). They obey $\epsilon_\pm=\epsilon^*_\mp$, and both have the same non-vanishing negative module with $\bar{\epsilon}_\pm \epsilon_\pm = -1$. Of course, we also obtain orthogonality, $\bar{\epsilon}_\pm \epsilon_ \mp = 0$. Moreover, $p_\mu \epsilon^\mu_\pm =0$. These two transverse polarization vectors have no time component and therefore are also eigenvectors of the helicity based on the matrix operators $\mathbf{L}^\pm$. The corresponding definition is
\begin{equation}
	\label{eq:13a}
	H^\pm(\hat{\mathbf{p}}) = \mathrm{i} \left(
	\begin{array}{cccc}
		0  & \mp x & \mp y & \mp z \\
		\mp x & 0 & -z & y \\
		\mp y & z & 0 & -x \\
		\mp z & -y & x & 0 \\
	\end{array}
	\right). 
\end{equation}
The tranverse polarization four-vectors obey $H^+(\hat{\mathbf{p}}) \epsilon_\pm = \pm \epsilon_\pm$, as well as $H^-(\hat{\mathbf{p}}) \epsilon_\pm = \pm \epsilon_\pm$. And therefore the dyadic matrix operator of the spatial second-order derivative in the wave equation (\ref{eq:7}) gives after Fourier transformation $H^\pm(\hat{\mathbf{p}})\,H^\mp(\hat{\mathbf{p}})\,\epsilon_\pm=\epsilon_\pm$. Thus in Fourier space the vector-field can be expressed by a superposition of these two transverse polarizations. 

However, in addition we also have to consider the longitudinal mode $\bm{\chi}_0$ for a massive spin-one particle \cite{wigner,bargman}. We find indeed that $\epsilon^\mu_0=(1,\bm{\chi}_0)$ could be a related relativistic eigenvector, obeying $H^\pm(\hat{\mathbf{p}})\epsilon_0
= \pm \mathrm{i}\epsilon_0$, and thus we have $H^+(\hat{\mathbf{p}})\,H^-(\hat{\mathbf{p}})\,\epsilon_0=\epsilon_0$.
But it has a module of zero, $\bar{\epsilon}_0\epsilon_0 = 0$, and thus is unphysical. But when following conventional wisdom \cite{schwartz}, 
a meaningful longitudinal polarization four-vector can be constructed as $\epsilon^\mu_0=(p,E(p)\bm{\chi}_0)/M$, with $E(p)=\sqrt{M^2+p^2}$. However, it is not an eigenfunction of $H^\pm(\hat{\mathbf{p}})$, but trivially of $H^\mp(\hat{\mathbf{p}})H^\pm(\hat{\mathbf{p}})=\sf{1}_4$. Like the transverse polarization vectors this longitudinal one also obeys $p_\mu \epsilon^\mu_0 =0$, with $p_\mu=(E(p), -p\bm{\chi}_0)$. These three polarization vectors are orthonormal, and they all have a negative module of $-1$. Moreover, they are valid for any spatial orientation of the momentum three-vector $\mathbf{p}$.

They can form a complete basis in Minkowski space when being added by the polarization vector associated formally with the scalar field. A possible natural choice is the four-momentum vector itself, which as shown above is perpendicular to the three other polarizations. We may thus adequately chose $\epsilon^\mu_s=p^\mu/M$, which in the boson rest frame is equal to $(1,0,0,0)^T$ at $\mathbf{p}=0$, and it yields a module of $\bar{\epsilon}_s \epsilon_s =1$. An enlightening full discussion of the issue of polarizations of the four-vector field can, when starting from the Proca (i.e., Maxwell with mass) equations, be found in the text book of Schwartz \cite{schwartz}, where also the role that the so-called little group of the LG plays in this context is elucidated. A comprehensive experimental review on how the W-boson polarization vectors can be measured at CERN is found in the article by Ballestrero et al. \cite{ballestrero}. 

Concerning the role of energetic stability of the four-vector quantum field $\phi^\mu=(\phi_0, \bm{\phi})$ to be discussed below, it is important to recapitulate that the modules of the associated four polarization vectors have the values $(1,-1,-1,-1)$, which corresponds to the trace of the space-time metric $g^{\mu\,\nu}$. This fact ensures that the energy density associated with the Lagrangian (\ref{eq:260}) remains positive definite after quantization.

  %%
  %% sec. 6, current density and stability
  %%
  \section{Current density and energetic stability of the vector quantum field }\label{sec6}
  
  In this section we provide first the rather general solutions of the kinetic equation of the four-vector vector quantum field that obeys equations (\ref{eq:7}) and (\ref{eq:75}), and then we discuss the related Noether current density. Concerning the spacetime dependence of $\phi(t,\mathbf{x})$ we make the standard ansatz of an exponential plane-wave function with the Lorentz-invariant phase, $\varphi(t,\mathbf{x})=E(p)t-\mathbf{p}\cdot\mathbf{x}$, which are the solutions of the second-order d'Alembert operator. As shown in the previous section, the field is characterized by its polarization vectors. We shall consider here a charged vector field and include the two transverse vectors $\epsilon_\pm(\hat{\mathbf{p}})$ as well as the longitudinal $\epsilon_0(\hat{\mathbf{p}})$ one. Consequently, we obtain the general field by summation over the momentum and helicity variables. This gives
  \begin{equation}
  	\label{eq:14}
  	\phi(t,\mathbf{x}) = 
  	\sum_{\mathbf{p},\alpha} 
  	\left[
  	\epsilon_\alpha(\hat{\mathbf{p}}) a_\alpha(\mathbf{p}) \exp{(-\mathrm{i}\varphi(t,\mathbf{x}))} + \epsilon^*_\alpha(\hat{\mathbf{p}})
  	b^\dagger_\alpha(\mathbf{p}) \exp{(\mathrm{i}\varphi(t,\mathbf{x}))}
  	\right],
  \end{equation}
  whereby we introduced the complex amplitudes indicated by $a$ for the particle (with negative frequency) and the hermitian-conjugated $b^\dagger$ for the antiparticle (with positive frequency), which both depend on the momentum $\mathbf{p}$ and helicity $\alpha$. They are the annihilation and creation operators in quantum field theory. We assume the vector-field to be charged, and thus $\phi$ becomes complex. Yet, if it is not charged, we can set $b^\dagger=a^\dagger$. Then the particle will be its own antiparticle. Let us consider just one of the plane-wave solutions in (\ref{eq:14}) with given momentum $\mathbf{p}$ and helicity $\alpha$:
  \begin{equation}
  	\label{eq:16}
  	\begin{array}{c}
  		\phi(\alpha,\mathbf{p}) = 
  		\epsilon_\alpha(\hat{\mathbf{p}})
  		a_\alpha(\mathbf{p}) \exp{(-\mathrm{i}\varphi(t,\mathbf{x}))} + 
  		\epsilon^*_\alpha(\hat{\mathbf{p}}) b^\dagger_\alpha(\mathbf{p}) \exp{(\mathrm{i}\varphi(t,\mathbf{x}))}, \\
  		\\
  		\bar{\phi}(\alpha,\mathbf{p}) = 
  		\bar{\epsilon}_\alpha(\hat{\mathbf{p}}) a^\dagger_\alpha(\mathbf{p}) \exp{(\mathrm{i}\varphi(t,\mathbf{x}))} + 
  		\bar{\epsilon}^*_\alpha (\hat{\mathbf{p}}) b_\alpha(\mathbf{p}) \exp{(-\mathrm{i}\varphi(t,\mathbf{x}))}.
  	\end{array}
  \end{equation}
  In order to be able to evaluate the current density we need the spatial and temporal derivative. The result is
  \begin{equation}
  	\label{eq:17}
  	\begin{array}{c}
  		\mathrm{i}\partial_\mu \phi(\alpha,\mathbf{p}) = (E(p), -\mathbf{p}) 
  		\left[\epsilon_\alpha(\hat{\mathbf{p}}) 
  		a_\alpha(\mathbf{p}) \exp{(-\mathrm{i}\varphi(t,\mathbf{x}))} - 
  		\epsilon^*_\alpha(\hat{\mathbf{p}}) b^\dagger_\alpha(\mathbf{p}) \exp{(\mathrm{i}\varphi(t,\mathbf{x}))}
  		\right], \\
  		\\
  		-\mathrm{i}\partial_\mu \bar{\phi}(\alpha,\mathbf{p}) =  (E(p), -\mathbf{p})
  		\left[\bar{\epsilon}_\alpha(\hat{\mathbf{p}})  
  		a^\dagger_\alpha(\mathbf{p}) \exp{(\mathrm{i}\varphi(t,\mathbf{x}))} - 
  		\bar{\epsilon}^*_\alpha(\hat{\mathbf{p}}) b_\alpha(\mathbf{p}) \exp{(-\mathrm{i}\varphi(t,\mathbf{x}))}
  		\right].
  	\end{array}
  \end{equation}
  We recall that $\epsilon_\pm(\hat{\mathbf{p}})=\epsilon^*_\mp(\hat{\mathbf{p}})$, and therefore particle and antipartice have opposite helicities for the same momentum $\mathbf{p}$.
  
  \subsection{Current density}
  In the case of a charged vector field its Lagrangian possesses a continuous symmetry under changes of the phase, which leads to a conserved current according to Noether's theorem. Taking the difference between the equation of motion and its adjoint version given in (\ref{eq:75}) we obtain, after multiplication of those equations from the right and left with the corresponding fields, the result for the temporal and spatial components of the current four-vector $j^\mu =(j_0, \mathbf{j})$ as follows 
  \begin{equation}
  	\label{eq:15}
  	\begin{array}{c}
  		j_0 = \mathrm{i}\left(\bar{\phi}(\frac{\partial}{\partial t}\phi) - (\frac{\partial}{\partial t}\bar{\phi})\phi \right), \\
  		\\
  		\mathbf{j} = \mathrm{i} \left( (\frac{\partial}{\partial \mathbf{x}} \bar{\phi} \cdot  \underline{\underline{\mathrm{L}}}) \phi 
  		- \bar{\phi}\,(\underline{\underline{\mathrm{L}}} \cdot \frac{\partial}{\partial \mathbf{x}}\, \phi) \right),
  	\end{array}
  \end{equation}
  which obey the standard continuity equation $\partial_\mu j^\mu= 0$. 
  The imaginary unit was inserted to make the current densities hermitian. Then by its definition $(j^\mu)^\dagger = j^\mu$. If the vector boson carries an electric charge, for example, then this current is just the electromagnetic one.
  
  We can now insert the results (\ref{eq:16}) and (\ref{eq:17}) into the current density (\ref{eq:15}) and obtain for a single Fourier component the expressions
  \begin{equation}
  	\label{eq:18}
  	\begin{array}{c}
  		j_0(\alpha,\mathbf{p}) = -2 E(p) \left(a^\dagger_\alpha(\mathbf{p})a_\alpha(\mathbf{p}) - b_\alpha(\mathbf{p})b^\dagger_\alpha(\mathbf{p})\right), \\
  		\\
  		\mathbf{j}(\alpha,\mathbf{p}) = - 2 \mathbf{p} \left(a^\dagger_\alpha(\mathbf{p})a_\alpha(\mathbf{p}) - b_\alpha(\mathbf{p})b^\dagger_\alpha(\mathbf{p})\right).
  	\end{array}
  \end{equation}
  In order to obtain this we used the expectation value of the matrix $\mathbf{L}^\pm$ as obtained with the eigenvectors (\ref{eq:13}), which for $\alpha=\pm 1$ gives
  \begin{equation}
  	\label{eq:19}
  	\bar{\epsilon}_\alpha \,\mathbf{L}^\pm\,\epsilon_\alpha =  -\alpha \hat{\mathbf{p}}.
  \end{equation}
  This result is readily obtained, if one multiplies this equation from the right by $\hat{\mathbf{p}}$ and uses the fact that the epsilons are eigenfunctions of the helicity $H^\pm(\hat{\mathbf{p}})$ and are normalized to $-1$. Moreover, we calculate the component $\hat{\mathbf{p}} \cdot \mathbf{j}$ of the current, which leads in (\ref{eq:15}) to the products $H^\pm(\hat{\mathbf{p}})\,H^\mp(\hat{\mathbf{p}})={\sf{1}_4}$, of which trivially the three $\epsilon_\alpha(\hat{\mathbf{p}})$ with $\alpha=(-1,0,+1)$ are eigenvectors. We now exploit the fact that the particle velocity is given by
  \begin{equation}
  	\label{eq:20}
  	\mathbf{v}(\mathbf{p})= \frac{\partial E(p)}{\partial \mathbf{p} } = \frac{\mathbf{p}}{E(p)} .
  \end{equation}
  Therewith we can correct for the field-normalization factors \cite{schwartz} in front (\ref{eq:18}) by division through $2E(p)$. Assuming that the vector boson carries a charge $q$, we can multiply the modified expression (\ref{eq:18}) by $q$, in which one can absorb the minus sign in front. Thus we obtain finally the four-vector charge current density as
  \begin{equation}
  	\label{eq:21}
  	j^\mu (\alpha,\mathbf{p}) =  q(1, \mathbf{v}(\mathbf{p})) \left(a^\dagger_\alpha(\mathbf{p})a_\alpha(\mathbf{p}) - b_\alpha(\mathbf{p})b^\dagger_\alpha(\mathbf{p})\right).
  \end{equation}
  This charge current density must be proportional to the number operators of the particle and antiparticle, which means we require commutators for the creation and annihilation operators,
  \begin{equation}
  	\label{eq:22}
  	[\,b_\alpha(\mathbf{p}),b^\dagger_\alpha(\mathbf{p})\,] =
  	b_\alpha(\mathbf{p}) b^\dagger_\alpha(\mathbf{p}) - b^\dagger_\alpha(\mathbf{p}) b_\alpha(\mathbf{p}) = 1,
  \end{equation}
  by means of which we can revert the sequence of the operators of the antiparticle in (\ref{eq:21}). Of course, the same commutator applies to the particles as well. Similar calculations can be carried out for the energy density based on the zeroth component of the energy momentum tensor \cite{schwartz} of the vector field based on the Lagrangian that is quoted below. In conclusion, the charged vector quantum field (\ref{eq:14}) describes bosons.
  
  \subsection{Energetic stability}
  Now we adress briefly the question of energetic stability of the quantum three-vector field $\bm{\phi}$, which is obtained from the spatial component of the epression (\ref{eq:14}). The summation of the polarization index then extends over the spin-one quantum numbers $\alpha = -1, 0, +1$, and the polarization vectors are given by $\bm{\chi}_\pm(\hat{\mathbf{p}}$ and $\bm{\chi}_0(\hat{\mathbf{p}})$. Again, we just consider a single Fourier component that reads
  \begin{equation}
  	\label{eq:16a}
  	\begin{array}{c}
  		\bm{\phi}(\alpha,\mathbf{p}) = 
  		\bm{\chi}_\alpha(\hat{\mathbf{p}}) 
  		a_\alpha(\mathbf{p}) \exp{(-\mathrm{i}\varphi(t,\mathbf{x}))} + 
  		\bm{\chi}^*_\alpha(\hat{\mathbf{p}}) b^\dagger_\alpha(\mathbf{p}) \exp{(\mathrm{i}\varphi(t,\mathbf{x}))}, \\
  		\\
  		\bm{\phi}^\dagger(\alpha,\mathbf{p}) = 
  		\bm{\chi}^\dagger_\alpha(\hat{\mathbf{p}}) 
  		a^\dagger_\alpha(\mathbf{p}) \exp{(\mathrm{i}\varphi(t,\mathbf{x}))} + 
  		\bm{\chi}^T_\alpha(\hat{\mathbf{p}}) b_\alpha(\mathbf{p}) \exp{(-\mathrm{i}\varphi(t,\mathbf{x}))}.
  	\end{array}
  \end{equation}
  Energetic stability requires to evaluate the 00-component of the energy momentum tensor, which is defined by the Lagrangian density (\ref{eq:262}) and reads
%%
%% Eq. 41
%%
  \begin{equation}
  	\label{eq:2425a}
  	\mathcal{E}_3 
          = \frac{
              \partial \mathcal{L}_3
            }{
              \partial (\partial_t \bm{\phi})
            }
            \partial_t \bm{\phi}  
            + \frac{
               \partial \mathcal{L}_3
              }{
               \partial(\partial_t \bm{\phi}^\dagger)
              } 
             {\partial_t \bm{\phi}^\dagger} - \mathcal{L}_3.
\end{equation}
By putting in the derivatives we obtain the result

\begin{equation}
  \label{eq:2425b}
   \mathcal{E}_3  
      = \frac{\partial \bm{\phi}^\dagger}{\partial t} 
        \frac{\partial \bm{\phi}}{\partial t} 
        + \frac{\partial \bm{\phi}^\dagger}{\partial \mathbf{x}} 
        \cdot \frac{\partial \bm{\phi}^\dagger}{\partial \mathbf{x}}
         + M^2 \bm{\phi}^\dagger\bm{\phi}.
\end{equation}
By insertion of the Fourier components of the fields given in equation
(\ref{eq:16a}) into the above equation of the energy density we obtain the corresponding Fourier contribution
\begin{equation}
\label{eq:2425c}
\mathcal{E}_3(\alpha,\mathbf{p}) =  ((E(p))^2 + p^2 + M^2) (\left(a^\dagger_\alpha(\mathbf{p})a_\alpha(\mathbf{p}) + b_\alpha(\mathbf{p})b^\dagger_\alpha(\mathbf{p})\right).
\end{equation}
Making again the above arguments for the use of commutators, this can be written as
\begin{equation}
\label{eq:2425d}
\mathcal{E}_3(\alpha,\mathbf{p}) =  E(p) (\left(a^\dagger_\alpha(\mathbf{p})a_\alpha(\mathbf{p}) + b^\dagger_\alpha(\mathbf{p})_\alpha(\mathbf{p})\right),
\end{equation}
whereby a field-normalization factor of $2E(p)$ \cite{schwartz} has been canceled. So each Fourier mode of the vector field adds a positive energy quantum to the overall energy density of the field. Therefore the Hamiltonian of the quantum field is bounded from below, and thus our Langrangian $\mathcal{L}$ (\ref{eq:262}) gives a physically meaningful theory. As already discussed above in previous sections, this Lagrangian is practically equal to that of four scalar fields assembled into a four-vector field. But it has in the three-vector section the important spin term, yet which is identically zero if there is no external gauge field coupling through the covariant derivative.

In comparison to the Proca theory, we obtain at the outset this important new ingredient owing to the use of the LG generators including the boson spin explicitly, which results in an expression for the boson magnetic moment being proportional to its spin one.

%%
%% sec. 7, coupling to magnetic field
%%
\section{The coupling of the vector boson to an electromagnetic  magnetic field}\label{sec7}

Finally, we discuss the before mentioned coupling of the spin of the charged and massive vector-boson field to a magnetic field. This connection is commonly described for scalar and spinor fields by means of the covariant derivatives
\begin{equation}
	\label{eq:26}
	D_t = \frac{\partial}{\partial t}  + \mathrm{i}q A_0(t,\mathbf{x}), \;\;
	D_\mathbf{x} = \frac{\partial}{\partial \mathbf{x}}  - \mathrm{i}q \mathbf{A}(t,\mathbf{x}).
\end{equation}
We can them insert into the kinetic equation (\ref{eq:7}), whereby we are interested just in the change of the spatial second-order derivative on the right-hand side. To evaluate it we make use of the formula (\ref{eq:A12}) in the Appendix. Since $D_\mathbf{x}$ does not commute with itself, we have now to consider both chiral signs, which means that we get
\begin{equation}
	\label{eq:27}
	(\mathbf{L}^\pm \cdot \mathbf{D}_\mathbf{x})(\mathbf{L}^\mp \cdot \mathbf{D}_\mathbf{x})= (\mathbf{D}_\mathbf{x} \cdot \mathbf{D}_\mathbf{x}){\sf{1}}_4 + \mathrm{i} \bm{\Sigma}^\pm \cdot (\mathbf{D}_\mathbf{x}\times\mathbf{D}_\mathbf{x}).
\end{equation}
According to equations (\ref{eq:7}),(\ref{eq:74}), and(\ref{eq:75}) we obtain
\begin{equation}
	\label{eq:277}
	\underline{\underline{\mathrm{L}}} : (\mathbf{D}_\mathbf{x} \mathbf{D}_\mathbf{x}) =  (\mathbf{D}_\mathbf{x} \cdot \mathbf{D}_\mathbf{x}){\sf{1}}_4 + \mathrm{i}\mathbf{J} \cdot (\mathbf{D}_\mathbf{x} \times \mathbf{D}_\mathbf{x}).
\end{equation}
The cross product produces the curl of the three-vector potential $\mathbf{A}$, which gives the magnetic field, $\mathbf{B}= \frac{\partial}{\partial \mathbf{x}}\times \mathbf{A}$. Thus the spin magnetic-field coupling for the vector boson of spin-one acoording to (\ref{eq:777}) and (\ref{eq:263}) reads finally for the spatial three-vector components of $\phi$ as follows,
\begin{equation}
	\label{eq:28}
	\mathrm{i}\mathbf{S} \cdot \mathbf{B} = 
	\left(
	\begin{array}{ccc}
		0 & B_z & -B_y \\
		-B_z & 0 & B_x \\
		B_y & -B_x & 0 \\
	\end{array}
	\right). 
\end{equation}	
This magnetic coupling operator only acts on the spatial component $\bm{\phi}$ of the vector field $\phi$. The temporal scalar component $\phi_0$ has no spin and yields no coupling to the magnetic field. It is interesting to compare this result with the spin one-half spinor-field coupling to the magnetic field, which by use of the Pauli matrices reads
\begin{equation}
	\label{eq:29}
	\bm{\sigma} \cdot \mathbf{B} = 
	\left(
	\begin{array}{cc}
		B_z  &  B_x - \mathrm{i} B_y  \\
		B_x +\mathrm{i} B_y & -B_z \\
	\end{array}
	\right). 
\end{equation}	
For a field oriented only in the $z$-direction, we get a diagonal matrix and the known doublet line splitting for a spinor fermion. But we get a more complicated situation for the vector boson. To understand better the meaning of (\ref{eq:28}), let us consider the special geometric situation that momentum and magnetic field point in the same direction, $\mathbf{B}=B\hat{\mathbf{p}}=B(x,0,z)$. Then it turns out that $\mathbf{S} \cdot \mathbf{B}$ acts in the same way as the spin helicity (\ref{eq:9}) on the polarization vector $\bm{\chi}_\pm (\hat{\mathbf{p}})$ of (\ref{eq:12}), and thus we obtain the usal result $\mathrm{i}\mathbf{S} \cdot  \mathbf{B} \,\bm{\chi}_\pm = \pm B \, \bm{\chi}_\pm$.

We can now make use of (\ref{eq:26}) and (\ref{eq:27}) in the original kinetic equation (\ref{eq:7}) of the vector boson. Thus we obtain a relativistic equation that includes the helicity of the rotation operator and reads
\begin{equation}
	\label{eq.30} 
	\left[ (\frac{\partial}{\partial t} + \mathrm{i} q A_0)^2  - (\frac{\partial}{\partial \mathbf{x}} - \mathrm{i}q\mathbf{A})^2
	+ M^2  - q \mathrm{i}\mathbf{J} \cdot \mathbf{B}
	\right] \phi = 0.
\end{equation}

The transition to a non-relativistic Schr\"{o}dinger-Pauli-type equation is obtained in the limit of the kinetic energy being much smaller than $Mc^2$, or equivalently the length scales are much larger than the Compton wavelength, $\lambda_C = \hbar /(Mc)$. We introduce only in this paragraph the speed of light $c$ and the Planck constant $\hbar$ again. Let us separate the very rapid Darwin oscillations at high frequency, $\hbar\omega_D = Mc^2$, from the slow variations at frequencies corresponding to non-relativistic energies by letting, $\phi = \tilde{\phi} \exp( -\mathrm{i}\omega_D t)$, in which the envelope or amplitude
wavefunction, $\tilde{\phi}$, is slowly varying, i.e. we have $\partial_t\ln(\tilde{\phi}) \ll \omega_D$. With this decomposition and for a quasi-static electrostatic potential $A_0$ the lowest-order terms yield
\begin{equation}
	\label{eq.31}
	(D_t^2+M^2)\phi=(-2\mathrm{i}M)  
	\exp(-\mathrm{i}M t) D_t \tilde{\phi}.
\end{equation}
With the aid of this approximation a kind of Schr\"{o}dinger-Pauli equation \cite{pauli1927} for the related three-vector component is obtained, which includes the spin helicity and reads
\begin{equation}
	\label{eq.32}
	\mathrm{i}  \frac{\partial}{\partial t} \tilde{\bm{\phi}}
	= \left[ 
            \frac{1}{2M} 
              ( \mathrm{i} \frac{\partial}{\partial \mathbf{x}} 
                + q\mathbf{A}
              )^2 + q A_0 
	    -  \mu_V \mathrm{i}
              ( 
                \mathbf{S} \cdot \mathbf{B}
              ) 
         \right] \tilde{\bm{\phi}},
\end{equation}
in which the magneton of the vector boson, $\mu_V = q\hbar/(2Mc)$, appears in front of the spin coupling term. It looks like the magneton,  $\mu_S = e\hbar/(2mc)$, of the spin one-half fermion with charge $e$ and mass $m$. However, the magneton of the spin-one vector boson has a gyromagnetic factor $g=1$. The algebraic reason for this result is, although the Sigma matrices behave algebraically just like the Pauli matrices, one half of their sum gives $\mathbf{J}$. And thus the coupling is to the rotation of four-vectors. This result originates essentially from the basic $SO(3,1)$ representation of the Lorentz group for a vector field as discussed in the introduction. 

The gyromagnetic factor derived here differs from the previous one obtained by Marsch \cite{marsch2017}, who calculated for any spin $s$ the gyromagnetic factor simply by use of the standard three-vector spin algebra. But this approach seems to give an erroneous value with $g=1/2$ for spin one. Our present result appears to be correct, and as shown above is consistent with boson statistics and commutators for the creation and annihilation operators.

%%
%% sec. 8, conclusion
%%
\section{Summary and conclusion}\label{sec8}

In this paper we have asked ourselves and answered the research question whether it is possible to make the spin-one property of a vector boson to appear explicitly and prominently in its kinetic equation, like it is the case for the spin one-half of a fermion described by the Dirac equation. Because for the massless photon as described by the Maxwell equations its spin or polarization is not obvious and therefore requires additional mathematical analysis. 

It turns out from our study that the helicity operator based on the combination of the two Lorentz-group generators written in terms of $\mathbf{L}^\pm$ are the decisive quantities providing new physics. Therefore, the resulting kinetic equations and Lagrangian of a massive charged vector boson are derived here from scratch, and discussed from the novel point of view exploiting directly the Lorentz group generators. Thereby, the chiral symmetry of the Lorentz transformation is guiding our algebraic calculations, and we did not have electromagnetic waves or photons in mind. So, our model is not a simple extension of the real Proca equation but a genuine different approach to describe massive charged vector bosons of spin one. 

Unlike for the Maxwell or any other gauge-field equations, gauge invariance plays no role here. Our Lagrangian is not based on the square of the usual gauge-field tensor \cite{schwartz} and the Lorentz condition, which together permit to reduce the degrees of freedom in conventional gauge-field theory to two degrees that are related only with tranverse polarizations. Exploiting the SO(3,1) symmetry of the Lorentz transformation in Minkowski space, we always get a decomposition of the original four-vector field into a scalar (time component) field and a three-vector (spatial components) field, having spin one and existing in real space. In order to accommodate a charge, these fields have to become complex and then describe particles and oppositely charged antiparticles. In the non-relativistic limit our theory yields a Schr\"{o}dinger-Pauli equation and a magneton of the vector-boson with a g-factor of unity.

A possible example for the application of our approach may be given by the electrically charged and massive $W^\pm$ vector bosons. Although they are unstable and extremely short lived, they should carry during their lifetime an electric current themselves, which could give rise to transient electromagnetic waves. This possibility arises by consideration of the boson Noether current as the source for an electromagnetic field. Such an option is not included in the SM, but may perhaps find some attention and application by the associated high-energy-physics community.

\section{Acknowledgements}\label{sec9}
The authors would like to thank very much an unknown referee for rather cricital but constructive reading of and helpful comments on a previous version of this paper. His review stimulated a major rewriting of parts of the material, and thus it led to substantial clarifications of various important issues.
\newline
This research received no external funding. No data were used and no new data were created. The authors declare no conflict of interest.
\newline 
Key ideas, conceptualization and writing by EM; review and editing by YN. Both authors have read and agreed to the published version of the manuscript.

%%
%% appendix
%%
%\appendixtitles{yes} % Leave argument "no" if all appendix headings %stay EMPTY (then no dot is printed after "Appendix A"). If the %appendix sections contain a heading then change the argument to "yes".
%\appendixstart
%\appendix

\section{Appendix: Relevant matrices}\label{sec10}

\subsection{Lorentz group-generator matrices}

In this subsection of the Appendix we compile some of the relevant matrices of the key physical quantities. For the generators of the Lorentz group we have the following $4\times4$ matrices. The component matrices of the rotation three-vector $\mathbf{J}$ read 
\begin{equation}
	\label{eq:A1}
	J_\mathrm{x}= 
	\left(
	\begin{array}{cccc}
		0 & 0 & 0 & 0 \\
		0 & 0 & 0 & 0 \\
		0 & 0 & 0 & -\mathrm{i} \\
		0 & 0 & \mathrm{i} & 0 \\
	\end{array}
	\right),
	J_\mathrm{y}= 
	\left(
	\begin{array}{cccc}
		0 & 0 & 0 & 0 \\
		0 & 0 & 0 & \mathrm{i} \\
		0 & 0 & 0 & 0 \\
		0 & -\mathrm{i} & 0 & 0 \\
	\end{array}
	\right),
	J_\mathrm{z}= 
	\left(
	\begin{array}{cccc}
		0  & 0 & 0 & 0 \\
		0 & 0 & -\mathrm{i} & 0 \\
		0 & \mathrm{i} & 0 & 0 \\
		0 & 0 & 0 & 0 \\
	\end{array}
	\right). 
\end{equation}
The component matrices of the boost three-vector $\mathbf{K}$ are also quoted here
\begin{equation}
	\label{eq:A2}
	K_\mathrm{x}= 
	\left(
	\begin{array}{cccc}
		0 & -\mathrm{i} & 0 & 0 \\
		- \mathrm{i}  & 0 & 0 & 0 \\
		0 & 0 & 0 & 0 \\
		0 & 0 & 0 & 0 \\
	\end{array}
	\right),
	K_\mathrm{y}= 
	\left(
	\begin{array}{cccc}
		0 & 0 & -\mathrm{i} & 0 \\
		0 & 0 & 0 & 0 \\
		- \mathrm{i} & 0 & 0 & 0 \\
		0 & 0 & 0 & 0 \\
	\end{array}
	\right),
	K_\mathrm{z}= 
	\left(
	\begin{array}{cccc}
		0 & 0 & 0 & -\mathrm{i}  \\
		0 & 0 & 0 & 0 \\
		0 & 0 & 0 & 0 \\
		- \mathrm{i} & 0 & 0 & 0 \\
	\end{array}
	\right). 
\end{equation}
From them we obtain straightforwardly the absolute value of the rotation and boost operator matrices as
\begin{equation}
	\label{eq:A3}
	\mathbf{J}^2 = \mathrm{diag}[0,2,2,2], \;\;
	\mathbf{K}^2 = -\mathrm{diag}[3,1,1,1].
\end{equation}
Consequently, we find $\mathbf{J}^2 - \mathbf{K}^2 = 3 {\sf{1}}_4$. Then for any three-vector $\mathbf{V}=(x,y,z)$ one obtains
\begin{equation}
	\label{eq:A4}
	\mathbf{J} \cdot \mathbf{V}= 
	\mathrm{i} \left(
	\begin{array}{cccc}
		0 & 0 & 0 & 0 \\
		0 & 0 &-z & y \\
		0 & z & 0 &-x \\
		0 &-y & x & 0 \\
	\end{array}
	\right), \;\;
	(\mathbf{J} \cdot \mathbf{V})^2= 
	\left(
	\begin{array}{cccc}
		0 & 0 & 0 & 0 \\
		0 & y^2+z^2 & -xy & -xz \\
		0 & -xy & x^2 + z^2 & -yz \\
		0 & -xz & -yz & x^2 + y^2 \\
	\end{array}
	\right),
\end{equation}
and similarly one obtains
\begin{equation}
	\label{eq:A5}
	\mathbf{K} \cdot \mathbf{V}= 
	\mathrm{i} \left(
	\begin{array}{cccc}
		0 & x & y & z \\
		x & 0 & 0 & 0 \\
		y & 0 & 0 & 0 \\
		z & 0 & 0 & 0 \\
	\end{array}
	\right),\;\;
	(\mathbf{K} \cdot \mathbf{V})^2= 
	-\left(
	\begin{array}{cccc}
		x^2+y^2+z^2 & 0 & 0 & 0\\
		0 & x^2 & xy & xz \\
		0 & xy & y^2 & yz \\
		0 & xz & yz & z^2 \\
	\end{array}
	\right). 
\end{equation}
Therefore, one finds that $(\mathbf{K} \cdot \mathbf{V})( \mathbf{J} \cdot \mathbf{V}) = (\mathbf{J} \cdot \mathbf{V}) (\mathbf{K} \cdot \mathbf{V}) =0$, which means that these scalar products commute with each other. Moreover, we obtain the important relation
\begin{equation}
	\label{eq:A7}
	(\mathbf{J}\cdot\mathbf{V})^2 - (\mathbf{K}\cdot\mathbf{V})^2 = (x^2+y^2+z^2){\sf{1}}_4
	= \mathbf{V}^2{\sf{1}}_4,
\end{equation}
which plays a key role and will be exploited in the main sections. 

\subsection{Spin matrices}

The Sigma spin matrices are given by the definition	$\bm{\Sigma}^\pm = -(\mathbf{J} \pm \mathbf{K})\Delta $. The three purely imaginary components of Sigma read
\begin{equation}
	\label{eq:A8}
	\Sigma^\pm_\mathrm{x}= \mathrm{i} \left(
	\begin{array}{cccc}
		0 & \mp 1 & 0 & 0 \\
		\pm 1 & 0 & 0 & 0 \\
		0 & 0 & 0 & -1 \\
		0 & 0 & 1 & 0 \\
	\end{array}
	\right),
	\Sigma^\pm_\mathrm{y}= \mathrm{i} \left(
	\begin{array}{cccc}
		0 & 0 & \mp 1 & 0 \\
		0 & 0 & 0 & 1 \\
		\pm 1 & 0 & 0 & 0 \\
		0 & -1 & 0 & 0 \\
	\end{array}
	\right),
	\Sigma^\pm_\mathrm{z}= \mathrm{i} \left(
	\begin{array}{cccc}
		0  & 0 & 0 & \mp 1 \\
		0 & 0 &-1 & 0 \\
		0 & 1 & 0 & 0 \\
		\pm 1 & 0 & 0 & 0 \\
	\end{array}
	\right), 
\end{equation}	
with the mixed commutator $[\bm{\Sigma}^\pm, \bm{\Sigma}^\mp]=0$. The $\Delta$ matrix connects the two Sigma matrices through $\Delta \bm{\Sigma}^\pm = \bm{\Sigma}^\mp \Delta$. By complex conjugation of the Sigma matrices in Eq.\,(\ref{eq:A8}), we can see that they, as being purely imaginary, obey $(\bm{\Sigma}^\pm)^* = -\bm{\Sigma}^\pm$. Moreover, the Sigma matrices fulfill, like the Pauli \cite{pauli1927} matrices, a metric condition in real space, namely
\begin{equation}
	\label{eq:A10}
	\Sigma^\pm_j \Sigma^\pm_k + \Sigma^\pm_k \Sigma^\pm_j = 2 \delta_{j,k} \sf{1}_4.
\end{equation}
Thus, the Sigma component matrices squared give unity, and their sum is $(\bm{\Sigma}^\pm)^2 =3\,\sf{1}_4$. Of course the Sigma matrices obey the angular momentum algebra $\bm{\Sigma}^\pm \times  \bm{\Sigma}^\pm = 2\mathrm{i}\bm{\Sigma}^\pm$. For two three vectors $\mathbf{A}$ and $\mathbf{B}$ we obtain the relation
\begin{equation}
	\label{eq:A11}
	(\bm{\Sigma}^\pm \cdot \mathbf{A})(\bm{\Sigma}^\pm \cdot \mathbf{B})= (\mathbf{A} \cdot \mathbf{B}){\sf{1}}_4 + \mathrm{i} \bm{\Sigma}^\pm \cdot (\mathbf{A}\times\mathbf{B}).
\end{equation}
From this equation we obtain by using $\bm{\Sigma}^\pm= -\mathbf{L}^\pm\Delta $ the important connection
\begin{equation}
	\label{eq:A12}
	(\mathbf{L}^\pm \cdot \mathbf{A})(\mathbf{L}^\mp \cdot \mathbf{B})= (\mathbf{A} \cdot \mathbf{B}){\sf{1}}_4 + \mathrm{i} \bm{\Sigma}^\pm \cdot (\mathbf{A}\times\mathbf{B}),
\end{equation}
which links the L matrices with the spin Sigma matrices. For $\mathbf{A}=\mathbf{B}$, the vector cross product vanishes, and thus the spin term is zero as well.


\begin{thebibliography}{}
      
    \bibitem{jackson}
    J.D. Jackson, \textit{Classical Electrodynamics}. Wiley, New York (1975)
    
    \bibitem{schwartz}
    M.D. Schwartz, \textit{Quantum Field Theory and the Standard Model}. Cambridge University Press, Cambridge, UK (2014)
    
    \bibitem{dirac1928}
    P.A.M. Dirac, The quantum theory of the electron. Proc. Roy. Soc. Lond. Ser. A, Math. Phys. Sci. \textbf{117}, 610 (1928)
    doi:10.1098/rspa.1928.0023
    
    \bibitem{bertone2005}
    G. Bertone, D. Hooper, J. Silk, 
    Particle dark matter: Evidence, candidates and constraints. 
    Phys. Rep. \textit{405}, 279 (2005)
    https://doi.org/10.1016/j.physrep.2004.08.031
    
    \bibitem{bertone2018}
    G. Bertone, D. Hooper, 
    History of dark matter, Rev. Mod. Phys. \textit{90}, 045002 (2018)
    https://doi.org/10.1103/RevModPhys.90.045002
    
    \bibitem{gaul2026}
    K. Gaul, L. Cong, D. Budker,
    Constraints on new vector boson mediated electron-nucleus interactions from spectroscopy data of polar diatomic molecules.
    Phys. Rev. Lett.  \textit{136}, 181805 (2026)
    https://doi.org/10.1103/d19m-s856 
     
    \bibitem{yangmills} 
    C.N. Yang, R.L. Mills, Conservation of isotopic spin and isotopic gauge invariance. Phys. Rev. \textbf{96}, 191 (1954)
    
    \bibitem{higgs}
    P. Higgs, Broken symmetries and the masses of gauge bosons. Phys. Rev. Lett. {\bf 13}, 508 (1964)
    
    \bibitem{weinberg}
    S. Weinberg, A model of leptons, Phys. Rev. Lett. \textbf{19}, 1264 (1967) doi:10.1103/PhysRevLett.19.1264
    
    \bibitem{proca1936}
    A. Proca, Sur la th\'eorie ondulatoire des 'electrons positif and negatif, J. Phys. Radium \textit{7}, 347 (1936).
    https://doi.org/10.1051/jphysrad:0193600708034700
    
    \bibitem{proca1938}
    A. Proca, Th\'eorie non relativiste des particules \`a spin entier. 
    J Phys. Radium \textit{9}, 61 (1938)
    https://doi.org/10.1051/jphysrad:019380090206100
    
    \bibitem{proca1939}
    A. Proca, S. Goudsmit, 
    Sur la masse des particules \'el\'ementaires.
    J. Phys. Radium \textit{10}, 209 (1939)
    https://doi.org/10.1051/jphysrad:01939001005020900
    
    \bibitem{proca2006}
    D.N. Poenaru, A. Calboreanu,
    Alexandru Proca (1897-1955) and his equation of the massive vector boson field. Europhysics News \textbf{37, No. 5}, 24 (2006)
  
    \bibitem{wigner}
	E. Wigner, On Unitary Representations of the Inhomogeneous Lorentz Group, Annals of Mathematics, Second Series. \textbf{40, No. 1}, 149 (1939) doi:10.2307/1968551
	
	\bibitem{bargman}
	V. Bargman, E. Wigner, Group theoretical discussion of relativistic wave equations. Proc. N.A.S. \textbf{34}, 211 (1948)
	
	\bibitem{joos}
	H. Joos, Zur Darstellungstheorie der inhomogenen Lorentzgruppe als Grundlage quantenmechanischer Kinematik. Fortschritte der Physik \textbf{10}, 65 (1962).
	
	\bibitem{mana2023}
	E. Marsch, Y. Narita, A new route to Symmetries through the Extended Dirac equation. Symmetry \textbf{15}, 492 (2023).
	
	\bibitem{kaku}
	M. Kaku, \textit{Quantum Field Theory, A Modern Introduction}. 
	Oxford University Press: New York, NY, USA (1993).
	
	\bibitem{pauli1927}
	W. Pauli, Zur Quantenmechanik des magnetischen Elektrons. Z. Phys. \textbf{43}, 601 (1927).
	
	\bibitem{mana2026}
	E. Marsch, Y. Narita, Kinetics and Lagrangian of vector bosons and spinor fermions as derived from Lorentz-group invariants. Eur.Phys.J. Plus \textbf{141}, 158 (2026)
	https://doi.org/10.1140/epjp/s13360-026-07359-4
	
	\bibitem{mana2020}
	E. Marsch, Y. Narita, Dirac equation based on the vector representation of the Lorentz group. Eur. Phys. J. Plus \textbf{135}, 782 (2020)
		
	\bibitem{marsch2017}
	E. Marsch, Relativistic wave equation for a massive charged particle with arbitrary spin.  Eur. Phys. J. Plus \textbf{132}, 188 (2017) https://doi.org/10.1140/epjp/i2017-11460-6
		
	\bibitem{ballestrero}
	A. Ballestrero, A. Maina, G. Pelliccioli,
	W boson polarization in vector boson scattering at the LHC.
	\textit{JHEP 003} 170 (2018)
	https://doi.org/10.1007/JHEP03(2018)170
	
	

\end{thebibliography}
\end{document}